\documentclass[12pt]{article}

\newcommand{\be}{\begin{equation}}
\newcommand{\ee}{\end{equation}}
\newcommand{\ba}{\begin{eqnarray}}
\newcommand{\ea}{\end{eqnarray}}

\newcommand{\la}{\lambda}

\newcommand{\e}{\epsilon}

\begin{document}

\hsize36truepc\vsize51truepc
\hoffset=-.4truein\voffset=-0.5truein
\setlength{\textheight}{8.5 in}

\begin{titlepage}
\begin{center}
\hfill \\
\hfill August 24, 2001\\
\hfill {LPTENS-01/29}
\vskip 0.6 in
{\large {\bf{ TWIST FREE ENERGY}	}}
\vskip .6 in
       \begin{center}
{\bf E. Br\'ezin$^{a)}$}  {\it and} {\bf C. De Dominicis$^{b)}$}
\end{center}
\vskip 5mm
\begin{center}
{$^{a)}$ Laboratoire de Physique Th\'eorique, Ecole Normale
Sup\'erieure}\\ {24 rue Lhomond 75231, Paris Cedex 05,
France}{\footnote{
 Unit\'e Mixte de Recherche 8549 du Centre National de la
Recherche
Scientifique et de l'\'Ecole Normale Sup\'erieure.
 }}\\ {\it{brezin @corto.lpt.ens.fr}}\\
{$^{b)}$ Service de Physique Th\'eorique, CE Saclay,\\ 91191
Gif-sur-Yvette, France}\\
{\it{cirano@spht.saclay.cea.fr}}
\end{center}

 \vskip 0.5 cm
{\bf Abstract}
\end{center}
One may impose to a  system with spontaneous broken symmetry, boundary
conditions which correspond to different pure states at two ends of a
sample. For a discrete Ising-like broken symmetry, boundary conditions
with opposite spins in two parallel limiting planes, generate an
interface and a cost in free energy per unit area of the interface.
For continuum symmetries the order parameter interpolates smoothly between
the end planes carrying two different directions of the order
parameter. The cost in free energy is then proportional to $L^{d-2}$ for
a system of characteristic size L. The power of $L$ is related to the
lower critical dimension, and the coefficient of this additional free
energy vanishes at the critical temperature. In this note it is shown
within a loop expansion that one does find the expected behavior of this
twist  free energy. This is a preamble to the study of situations
where the broken continuum symmetry is believed to be more complex, as
in Parisi's ansatz for the Edwards-Anderson spin glass.

\vskip 14.5pt
\end{titlepage}
\setlength{\baselineskip}{1.5\baselineskip}


\section{Introduction}
Spontaneously broken symmetries are characterized by the
existence of several possible pure states. If one imposes
"twisted" boundary conditions , i.e. different pure states at two ends of
the system, the free energy per unit volume will be slightly greater than
the free energy corresponding to one single pure state over the whole
system.

 For
a simple discrete symmetry, such as the $Z_2$-symmetry of Ising-like
systems, one may consider an (hyper)-cubic system with up spins in the
$z=0$ plane, down spins in the $z=L$ plane and for instance periodic
boundary conditions in the transverse directions $x_1, x_2, \cdots
x_{d-1}$. This will generate an interface in the system centered around
some plane
$z=z_0$ and a cost in free energy
\be \label{Ising}\Delta F=  F_{\uparrow,\downarrow}-
F_{\uparrow,\uparrow} =
\sigma L^{d-1}
\ee in which the interfacial tension $\sigma (T)$ is finite at low
temperature, but vanishes at the critical temperature as
\begin{equation} \sigma(T) = \sigma_0 (\frac{T_c -T}{T_c})^\la.
\end{equation}
As is well-known  the power $(d-1)$ of $L$ in
(\ref{Ising}) implies that the lower critical dimension of systems with a
discrete symmetry is equal to one, i.e. there is no ordered phase unless
$d$ is greater than one. Widom
\cite {Widom} has first proposed the scaling law
\be \la = (d-1)\nu \ee
in which $\nu$ is the correction length exponent
\be \xi = \xi_0 (\frac{T_c -T}{T_c})^\nu.\ee
The corresponding amplitude relation implies that the combination
\be \xi_0^{d-1} \sigma_0 \ee
is universal. All this was studied long ago \cite {BF} by renormalization
group techniques and $(4-d)$-expansion. At leading order the classical
(mean field) solution is a kink, of hyperbolic tangent shape,
interpolating between up and down spins, and fluctuations are given at
one-loop order by the Fredholm determinant of a one-dimensional
Schrodinger operator in a $1/{\cosh^2(z-z_0)}$ potential which, as is
well-known, is solvable analytically.

For more complex spontaneously broken symmetries, continuum symmetries,
or replica-symmetry breaking, the situation is less trivial, and
it it is necessary to look into the problem in some detail in order
to understand the lower critical dimension. For a continuum symmetry
group G, broken down to a subgroup H, as in N-vector models, one
considers the free energy with two different pure states in the two
planes $z=0$ and $z=L$. For an N-vector model one considers for
instance an order parameter uniform along the vector $(1,0,\cdots,0)$
in the
$z=0$ plane, and uniform but rotated by an angle $\theta_0$ in the
plane $z=L$,  i.e. lying along the vector
$(\cos{\theta_0},\sin{\theta_0},0,\cdots,0)$. There again one expects
a cost in free energy
\be \Delta F = \sigma(T,\theta_0) L^{d-2} \ee
in agreement with a lower critical dimension equal to two, and with a
"twist" energy $\sigma(T,\theta_0)$ (or spin stiffness constant)
vanishing as
$\theta_0^2$ for small $\theta_0$,( the ratio $\sigma/\theta_0^2$ is the
helicity modulus defined by Fisher, Barber and Jasnow \cite{FBJ}), and
vanishing at
$T_c$
\cite{J, RJ} like
$(T_c-T)^{\nu(d-2)}$. If it is quite elementary to verify these
statements within mean field theory, not difficult also to check them in
the vicinity of the lower critical dimension $d_l=2$ through the
non-linear sigma model \cite{Ch, Elka}, it is not so simple to examine the
problem below the upper critical dimension
$d_u=4$. This note is thus devoted to this point. Our aim in
performing this calculation is to repeat it later for a spin glass . There,
below the temperature of transition, one recovers a broken
continuum symmetry, displayed by Parisi's ansatz \cite {Parisi} of
replica symmetry breaking for the Edwards-Anderson model,
which yields a continuum of schemes with equal free energy
(reparametrization invariance).  If the situation at $d_u=6$ is more or
less under control, despite its complexity \cite{DD}, the knowledge about
$d_l$ is poor, although it is believed to lie between two and three
\cite{DD,KT, Hil}. If one imposes again two different schemes at two ends
of the system, one expects \cite{FP} a cost in free energy
\be \Delta F = \sigma L^{d-2+\eta} \ee
with some negative anomaly $\eta$ which would yield a lower critical
dimension $d_l= 2-\eta$. The possible presence of this anomaly
requires to compute at least a one-loop correction to mean field
theory. This difficult calculation will be reported in a subsequent
article, and this note simply aims at showing that already for the
well understood N-vector model, the theory is  somewhat
involved. The rest of this note is thus  devoted to the N-vector
model below four dimensions, treated thus through a
$(\vec{\phi}^2)^2$ field theory and an $\epsilon = 4-d$ expansion.It is
interesting to note that a direct
calculation of the helicity modulus has also been performed directly in
three dimensions, in spite of the singularities expected from Goldstone
massless modes \cite {BSD, SMD}. This has been done by keeping a symmetry
breaking field until one can let it go safely to zero at the end  of the
calculation of this helicity modulus.

\section {Mean field theory}

The action for the N-vector model in the broken symmetry domain is
\be  \label{8} S = \int_0^{L} dz \int d^{d-1}x_{\perp}\  \big[
\frac{1}{2}(\nabla
\vec{\phi})^2 - \frac{1}{2} \vert t\vert( \vec{\phi})^2 +\frac{g}{4}
(\vec{\phi})^2)^2 \big], \ee
in which $t$ is proportional to $T-T_c$. A pure state throughout the
bulk would have a magnetization $\vec M$ whose magnitude is given by
\be \vert t\vert  = g M^2 .\ee Subtracting the bulk contribution one
thus has
\be \Delta S = \int_0^{L} dz \int d^{d-1}x_{\perp}  \big[
\frac{1}{2}(\nabla
\vec{\phi})^2 + \frac{g}{4}
(\vec{\phi}^2 -M^2)^2 \big]. \ee
The free energy $\Delta F$ is the value of the minimum of $\Delta S$
with the  boundary conditions
\ba \vec \phi (z=0, \vec x_{\perp}) &=& M(1,0, \cdots,0)\nonumber \\
\vec \phi (z=L, \vec x_{\perp}) &=& M(\cos{\theta_0},\sin{\theta_0},0,
\cdots,0).\ea
We fix here the value of the order parameter on the edges, rather
than imposing magnetic fields on the boundaries. Our partition function
will thus be defined with fixed prescribed values of the order parameter
on the two edges, rather than fixing a surface magnetic field and letting
the surface order order parameter fluctuate, as in the work of M. Krech
\cite{K} for instance. For an $N=1$ (scalar) order parameter, we would
have to fix the surface order parameter to a value slightly smaller than
the bulk magnetization, but for $N>1$ one can directly take the modulus
of the surface order parameter equal to the bulk magnetization, as
shown in the mean field solution of the equations of motion below. It is
easy to verify that
$\Delta S$ is minimum
\begin {itemize}
\item when the order
parameter remains in the 2-plane of the two vectors defined by the
boundary conditions
\item
when $\vec \phi $ is a function of z-alone, i.e. independent of
$x_\perp$.
\end {itemize}
and one can parametrize the mean field solution  as
\be \vec \phi = \rho(z) ( \cos \theta(z), \sin \theta (z), 0,
\cdots,0) , \ee
for which
\be \Delta S = L^{d-1} \int_0^{L} dz   \big[
\frac{1}{2}(\frac{d\rho}{dz})^2 + \frac{1}{2} \rho
^2(\frac{d\theta}{dz})^2 +\frac{g}{4} (\rho^2 -M^2)^2 \big]. \ee
The solution will be close to that of an order parameter
uniformly rotating between the two planes with a constant magnitude
$M$, namely $\rho(z) = M$ and $\displaystyle\theta (z) = \frac{z}{L}
\theta_0$, for which $\Delta F = \frac{1}{2} {\theta_0}^2M^2 L^{d-2}$.
However, although the solution is close to that for large L, we
shall need the corrections of order $1/L^2$ to that simple ansatz,
and one has to solve the variational equations
\ba \frac{d}{dz}(\rho^2 \theta^{\prime}) &=& 0 \nonumber
\\
\rho^{\prime\prime} - {\rho \theta^{\prime}}^2 - g \rho(\rho^2-M^2)&=&
0.\ea Defining the dimensionless variables
\ba \tau &=& z \sqrt{2gM^2} \nonumber \\
r(\tau) &=& \frac{1}{M}\rho(z), \ea
 the equations of motion are easily cast into the form
\ba \label{EM}(\frac{dr}{d\tau})^2 &=& \omega - v(r)\nonumber \\
r^2 \frac{d\theta}{d\tau} &=& \sqrt{\gamma} , \ea
with
\be v(r) = \frac{\gamma}{r^2} - \frac{1}{4}( 1-r^2)^2 .\ee
We can think of the equation for $r$ as an equation of motion in
$(r,\tau)$-plane in which $r$ starts at $r=1$ for $\tau=0$ ,
decreases down to some $r_0$ , then increases and returns to $r=1$ at
\be \tau_0 = L \sqrt{2gM^2}\  .\ee
The  parameters $\gamma$ and $\omega$ have still to be determined by
the boundary conditions. Of course the exact solution of the
equations of motion (\ref{EM}) involves elliptic functions \cite{K}.
However it turns out that      it is sufficient for our purpose to
consider the regime in which
$\gamma$ is small, which corresponds to $L$ large
compared to the correlation length $\xi$ (or if $L/\xi$ is finite,
corresponds to small $\theta_0$). Indeed in that regime the order
parameter has essentially a fixed length $r(\tau)$ close to 1, and
\ba \label{19}\frac{\theta_0^2}{2 g M^2 L^2} &=&\gamma
+O(\gamma^2)\nonumber\\
\omega &=& \gamma+ O(\gamma^2)\nonumber \\
r &=& 1 -\gamma s(\tau) + O(\gamma^2). \ea
The full integration  to this order
in $\displaystyle \frac{\theta_0^2}{2 g M^2 L^2}$is then easy and
leads to
\be \label {20} s(\tau) = 1 - \frac {\cosh \vert
\frac{\tau_0}{2}-\tau\vert} {\cosh (\frac{\tau_0}{2})}\  .\ee
To that same order one finds
\be \label {MF}\Delta F = \frac{1}{2} \theta_0^2 M^2 L^{d-2}
+O(\gamma^2) =
\frac{1}{2g} \theta_0^2  L^{d-2} \vert t \vert +O(\gamma^2). \ee
Let us note that,  in mean field, the correlation length is related
to the temperature by
\be \xi^{-2} = 2 g M^2 = 2\vert t\vert, \ee
and thus $\gamma$ is small either because $L/\xi$ is large or because
$\theta_0$ is small. The result (\ref{MF}) is thus in agreement with
our expectations
\be  \Delta F = \sigma L^{d-2}\ee
with $\sigma = \frac{1}{2g} \theta_0^2  \vert t \vert $ vanishing
at the critical temperature . We also verify the scaling law  $\sigma
(t)\sim
\vert t\vert^ {\nu(d-2)}$ (which is expected to be true for $d\leq
4$) in four dimensions at which
$\nu = 1/2$ and
$\nu(d-2) =1$.

As far as mean field theory is concerned the picture is simple : for
$\theta_0 \xi/L$ small, the magnitude of the order parameter remains
close to M over the whole sample, and its direction smoothly
interpolates between the two end planes with a constant angle
gradient. If we went beyond this simple picture in (\ref{20})
it is because this will be needed in the loop expansion when we
consider fluctuations around the mean field.
\section {One loop corrections}
We now go to dimension $d= 4-\epsilon$ and work to first order in
$\epsilon$, which requires the calculation of one-loop fluctuations
around mean field theory. Instead of an ultra-violet cut-off given
by some lattice spacing, it turns out to be much more convenient, as
often, to use dimensional regularization.
The mean field solution is
\be \vec \phi_c = M r(z) (\cos{ \theta(z)}, \sin {\theta(z)}, 0,
\cdots,0), \ee
with $r$ and $\theta$ described in the previous section. It is
convenient to introduce an orthonormal moving frame consisting of the
vectors
\ba \vec e_1 &=& (\cos{ \theta(z)}, \sin {\theta(z)}, 0,
\cdots,0) \nonumber \\ \vec e_2 &=& (-\sin{ \theta(z)}, \cos
{\theta(z)}, 0,\cdots,0),\ea
plus the $(N-2)$ fixed unit vectors $\vec e_a, (a=3,\cdots,N)$
perpendicular to the two-plane (1-2). The field $\vec
\phi(z, \vec {x_\perp})$ is then parametrized as
\be \vec
\phi(z, \vec {x_\perp}) = (\rho(z) + \psi_1(z, \vec {x_\perp}))\vec
e_1(z) + \psi_2(z, \vec {x_\perp})\vec e_2(z) + \sum_{a=3}^N
\psi_a(z, {x_\perp})\vec e_a . \ee
The boundary conditions on those $\psi_a$ are periodic in the
tranverse directions and, since the mean field order parameter $phi_c$ is
equal to the magnetization on the boundaries, one has to impose
Dirichlet conditions on the fluctuating fields
$\psi_a(z=0) =
\psi_a(z=L)=0$. A one-loop calculation requires to keep only the
quadratic terms in
$\psi_a$ of the action. Collecting those terms one finds
\be S = S_0 + S_2 \ee
in which $S_0$ is the mean field contribution and
\ba S_2 = &&\int_0^L dz \int d^{d-1}x_{\perp}\big[ \frac{1}{2}\sum_1^N
(\nabla \psi_a)^2 + \frac{1}{2}(2gM^2 + (\frac{d\theta}{dz})^2 +
3gM^2(r^2(z)-1))\psi_1^2 \nonumber \\+&& \frac{d\theta}{dz}(\psi_1
\frac{\partial
\psi_2}{\partial z}-\psi_2 \frac{\partial
\psi_1}{\partial z}) +\frac{1}{2}((\frac{d\theta}{dz})^2
+gM^2(r^2(z)-1))\psi_2^2
\nonumber\\
+&&\frac{1}{2}gM^2(r^2(z)-1)\sum_3^N \psi_a^2
\big] .\ea

The one-loop free energy is thus equal to the properly normalized
\be \Delta F = S_0 + \frac{1}{2}\rm{Tr} \ln \frac{\partial^2
S_2}{\partial
\psi_a(x)\partial\psi_b(y)} .\ee
The normalization will be chosen such that $\Delta F$ vanishes with
$\theta_0$. \\
{\bf{Contribution of the N fluctuating modes}}
\begin{itemize}

\item The transverse modes $\psi_a$, $a=3,\cdots, N$
are decoupled and give a contribution to $\Delta F$ equal to
\ba \label{transverse}&&\frac{1}{2}(N-2) \rm{Tr}\ln[ -\nabla^2 +
gM^2(r^2(z)-1) ]\nonumber\\&&=
\frac{1}{2}(N-2)L^{d-1}\int
\frac{d^{d-1}q_{\perp}}{(2\pi)^{d-1}}
\rm{Tr}\ln[q_\perp^2 -\frac{d^2}{dz^2} + gM^2(r^2(z)-1) ].\ea
On should note that although $r^2(z)-1$ is negative, the spectrum
of $  -\frac{d^2}{dz^2}$ is bounded below by $\pi^2/L^2$ since we
have  Dirichlet boundary conditions on the planes $z=0$ and
$z=L$.  Taking  the explicit solution (\ref{19},\ref{20}) one sees
that the spectrum of  $-\frac{d^2}{dz^2} + gM^2(r^2(z)-1) $ is
bounded below by $(\pi^2-\theta_0^2)/L^2$ and is thus positive.

Therefore a priori one has to compute the Fredholm determinant of
a one-dimensional Schrodinger operator in the complicated potential
$r^2(z)$. However for large $L$, perturbation theory gives very simply
the answer since $r^2(z)-1$ is of order $1/L^2$. This is to be
contrasted with a localized Ising interface, for which there is no
small parameter for large $L$. The simplification here is due to
the fact  that the order parameter turns slowly from one end of the
system to the other one and thus has only small fluctuations in the
moving frame that we have introduced.

Then we may replace $\rm{Tr}\ln[q_\perp^2 -\frac{d^2}{dz^2} +
gM^2(r^2(z)-1) ]$ (subtracted to vanish at $\theta_0 = 0$) by
$gM^2\rm{Tr}[q_\perp^2 -\frac{d^2}{dz^2}]^{-1} (r^2(z)-1) $.
Expanding on the basis of the Dirichlet eigenstates of $
-\frac{d^2}{dz^2}$, the states $\sqrt{\frac{2}{L}}\sin {(n\pi z/L)}$,
we obtain the contribution of these modes, in the large $L$ limit,  under
the form
$$ -(N-2)\theta_0^2 L^{d-3} \int
\frac{d^{d-1}q_{\perp}}{(2\pi)^{d-1}}\frac{1}{L}\int_0^L dz
\ s(z)\sum_{n=1}^{\infty}
\frac{\sin^2{(n\pi z/L)}}{q_{\perp}^2+ (n\pi/L)^2}.
$$
in which $s(z)$ is the explicit mean field correction (\ref{20}). In the
large L limit one can replace
$s(z)$ by one, the sum over
$n$ by an integral which, combined with the integral over $q_{\perp}$,
gives the integral
$L\int d^d p/p^2$ which vanishes in dimensional regularization. Those
modes have thus a vanishing contribution to the terms proportional to
$L^{d-2}$ of $\Delta F$.
\item
We now come to the coupled $\psi_1$-$\psi_2$ modes, using again that
$r^2(z)-1$ and $(d\theta/dz)^2$ are of order $1/L^2$. This allows
one again to use a perturbation expansion about a massless
$\psi_2$-mode and a massive
$\psi_1$. After a lengthy, but elementary calculation, we obtain the
contribution of these two modes to
$\Delta F$ under the form of a sum of five terms :
\ba \label{31}\Delta
F^{\rm{one-loop}} =&& \frac{\theta_0^2}{2L^2}\rm{Tr}(\frac{1}{-\nabla^2 +
2gM^2}+\frac{1}{-\nabla^2 } )\nonumber \\ &&-
\frac{\theta_0^2}{2L^2}\rm{Tr}(3\frac{1}{-\nabla^2 +
2gM^2}s(z)+\frac{1}{-\nabla^2 }s(z) ) \nonumber \\
&&-2\frac{\theta_0^2}{2L^2}\rm{Tr}\large(\frac{1}{(-\nabla^2 +
2gM^2)(-\nabla^2) }(-\frac{\partial^2}{\partial z^2} )\large) . \ea
We leave the detail of the calculations to an appendix and simply report
the result. We have computed the $1/\e$ pole of this expression, for
arbitrary $L/\xi$ and obtained.
\be \label{32} \Delta F =  \frac{\theta_0^2}{2g}L^{d-2} \vert t \vert +
\frac {3}{8\pi^2 \e}\theta_0^2 L^{d-2} \vert t \vert ^{1-\e/2} \ee
(we have kept it under this form since $g$ and $\vert t \vert ^{\e/2}$
have the same dimension). In this expression we have kept the finite $
(\vert t\vert\ln \vert t \vert)$  term and neglected the non-logarithmic
terms.
\end{itemize}
\section {Renormalization and scaling}
We first note that the pole in $1/\e$ in (\ref{32}) is independent of
$L/\xi$, as it should, since the renormalizations are independent of
this ratio. Next we note that the limit of $\e$ going to zero should be
finite, provided we perform a coupling constant and mass renormalization
(there is no wave function renormalization at this one-loop order).
Taking the standard one loop result from literature \cite{ZJ} ( with the
appropriate normlization of the coupling constant chosen in (\ref{8}))
one has, at one-loop,
\be \frac{1}{g} = \mu^{-\e}(\frac{1}{g_R}  - \frac{N+8}{8\pi^2 \e}) \ee
for the coupling constant renormalization ($\mu$ is an arbitrary inverse
length scale) and
\be t = t_R (1 + g_R \frac{N+2}{8\pi^2 \e})\ee
for the mass (i.e. temperature) renormalization.
This gives a renormalized  expression for $\Delta F$ in terms of $g_R$
and $t_R$ which is finite, as expected, when $\e$ goes to zero:
\be \label{33} \Delta F =  \frac{\theta_0^2}{2g_R}L^{d-2}\mu^{-\e} \vert
t_R\vert ( 1 - \frac {3g_R}{8\pi^2} \ln{ \frac{\vert t_R\vert }{\mu^2}}).
\ee

The scaling of the coefficient of $\theta_0^2 L^{d-2}$ in the critical
region requires a replacement of $g_R$ by the infra-red stable fixed
point
\be g_R ^* = \frac {8\pi^2 \e}{N+8} + O(\e^2) , \ee
and the exponentiation
\be ( 1 - \frac {3g_R}{8\pi^2} \ln{
\frac{\vert t_R\vert }{\mu^2}}
) \to (\frac{\vert t_R\vert }{\mu^2})^{3g_R^*/8\pi^2}.
 \ee
Given that the correlation legth exponent $\nu$ has the expansion
\be \nu = \frac{1}{2} +  \frac{N+2}{4(N+8)} \e + O(\e^2) ,\ee
one verifies to this order that
\be 1 - \frac{3g_R^*}{8\pi^2}  = \nu(d-2)\ee
which does yield the expected scaling law for the vanishing of the twist
energy at $T_c$ in the $O(N)$-model.

\section{ Final remarks}

Although a priori more cumbersome than the calculation of the
interfacial energy for a discrete symmetry, it turns out that, for a
continuum symmetry, it is possible to compute the complicated Fredholm
determinant of fluctuations around mean field theory by  an expansion in
powers of
$1/L$ which was not available for an interfacial wall. The calculation
involves a description of the mean field solution in which it is not
sufficient to simply assume that the order parameter rotates
 with  a
constant angle gradient from end to end, with a fixed length
equal to the magnetization. The calculation presented here may be easily
generalized to any continuum symmetry group G, broken down
spontaneously below a critical temperature to a subgroup H, with an order
parameter in a given irreducible representation R of G.

This calculation provides  an explicit test  of the fact that
 the renormalizations are the same around any background
solution: in the usual case one expands about a classical solution which
is constant over the sample,   whereas here one
expanded around a non-trivial solution, and yet we found that the same
coupling constant and mass renormalizations did work. We have also
verified that the finiteness of the end result of the free energy for
any ratio
$L/\xi$. However the method that we have followed, has  made use of a
small parameter, namely
$\theta_0\xi/L$. Away from the critical temperature this parameter is
small because $L$ is large. However if
$L/\xi$ is finite our calculation is restricted to small $\theta_0$. In
the finite $L/\xi$ regime, $\Delta F$ is a priori a more complicated
function of $\theta_0$ for which we have only determined the first
term. Let us stress also that we have used the $\epsilon$-expansion,
since we wanted to determine  the behavior of the twist free energy near
the upper critical dimension.
\newpage
\setcounter{equation}{0}
{\bf Appendix : {One-loop divergences}}
\vskip 5mm
Let us return to the five terms contained in (\ref{31}) for  the one-loop
calculation of $\Delta F$.
\begin {itemize}
\item
\be (a) = \frac{\theta_0^2}{2L^2}\rm{Tr}(\frac{1}{-\nabla^2 +
2gM^2}).\ee
If $L$ goes to infinity first we may simply neglect the quantization of
the longitudinal modes  and write
\be (a) = \frac{\theta_0^2}{2}L{^{d-2}}\int
\frac{d^dp}{(2\pi)^d}\ \frac{1}{p^2 + 2gM^2} = - \frac{1}{16 \pi^2 \e}
\theta_0^2  L^{d-2} (2gM^2)^{1-\e/2} \ee
in which it is understood that we have neglected the terms of order
$\e^0$.
For finite $L/\xi$ the calculation is much more involved. Going to the
large L limit for the tranverse periodic directions, but keeping the
quantization of the longitudinal modes one has
\ba &&(a) = \frac{\theta_0^2}{2}L{^{d-3}}\int
\frac{d^{d-1}q_{\perp}}{(2\pi)^{d-1}}\sum_{n=1}^{\infty} \frac{1}{q^2 +
\frac{n^2\pi^2}{L^2}+ 2gM^2}\nonumber \\ =
&&\frac{\theta_0^2}{8\pi}L{^{d-2}}\int_0^{\infty} dq
\frac{q^{d-2}}{\sqrt{q^2 +2gM^2}}[ \coth {L\sqrt{q^2
+2gM^2}} - \frac{1}{L\sqrt{q^2
+2gM^2}}]\nonumber \\
 =&&\frac{\theta_0^2}{8\pi}L{^{d-2}}(2gM^2)^{1-\e/2}\int_0^{\infty} dx
\frac{x^{d-2}}{\sqrt{x^2 +1}}[ \coth {l\sqrt{x^2
+1}} - \frac{1}{l\sqrt{x^2
+1}}], \nonumber \\\ea
in which
\be l = L\sqrt{2gM^2} = L/\xi. \ee
We now use the identity
\ba\label {5} &&\int_0^{\infty} dx
\frac{x^{d-2}}{\sqrt{x^2 +1}}[ \coth {l\sqrt{x^2
+1}} - \frac{1}{l\sqrt{x^2
+1}}] \nonumber \\
&&= \int_0^{\infty} dx
\frac{x^{d-2}}{\sqrt{x^2 +1}}[ 1 - \frac{1}{lx}]\nonumber \\
&&+ \int_0^{\infty} dx
\frac{x^{d-2}}{\sqrt{x^2 +1}}[ \coth {l\sqrt{x^2
+1}} - \frac{1}{l\sqrt{x^2
+1}}-1 + \frac{1}{lx}].\ea
The first term of the r.h.s. of (\ref{5}) is elementary and gives
$-1/(2\e)$ plus finite terms. It is easy to see that the second integral
of the r.h.s. of (\ref{5}) is finite when $d\to 4$.  This proves that
the divergent part of (a) is as expected independent of $l= L/\xi$.
\item
\be (b) = \frac{\theta_0^2}{2L^2}\rm{Tr}(\frac{1}{-\nabla^2 }).\ee
Again if $L$ goes to infinity first
\be (b) \to \frac{\theta_0^2}{2(2\pi)^{d}}L^{d-2}\int \frac{d^dp}{p^2}
\ee which vanishes in the dimensional regularization scheme.
\item
\ba (c) &=& -  \frac{3\theta_0^2}{2L^2}\rm{Tr}(\frac{1}{-\nabla^2+ 2gM^2
}s(z)) \nonumber \\&=& -
\frac{3\theta_0^2}{2L^2}\frac{L^{d-1}}{(2\pi)^{d-1}}\int
d^{d-1} q_{\perp} \frac {2}{L}\int_0^L dz
\sum_1^{\infty}(\frac{\sin^2{n\pi z/L}}{q_{\perp}^2 + n^2\pi^2/L^2 +
2gM^2 }s(z)). \ea
Again if one lets L go to infinity first one can replace $s(z)$ by one,
the calculations are then elementary and yield
\be (c) = \frac{3}{16\pi^2\e}(2gM^2)^{1-\e/2} \theta_0^2 L^{d-2}.\ee
For finite $L/\xi$ one can prove with the help of the explicit form for
$s(z)$ that the divergent part is unchanged.
\item
\be (d) = -  \frac{\theta_0^2}{2L^2}\rm{Tr}(\frac{1}{-\nabla^2}s(z))\ee
Again it is easy with the same integral representation
\be (d) = -
\frac{\theta_0^2}{2L^2}\frac{L^{d-1}}{(2\pi)^{d-1}}\int
d^{d-1} q_{\perp} \frac {2}{L}\int_0^L dz
\sum_1^{\infty}(\frac{\sin^2{n\pi z/L}}
{q_{\perp}^2 + n^2\pi^2/L^2 }s(z)), \ee
to prove that the leading term, proportional to $L^{d-2}$, multiplies
the integral $\int d^dp/p^2$ which vanishes.
Therefore
\be (d)=0\ee
\item
\ba (e) &=& -2\frac{\theta_0^2}{2L^2}\rm{Tr}(\frac{1}{(-\nabla^2 +
2gM^2)(-\nabla^2) }(-\frac{\partial^2}{\partial z^2} )) \nonumber\\
&=&  -
\frac{2\theta_0^2}{L^2}\frac{L^{d-1}}{(2\pi)^{d-1}}\int
d^{d-1} q_{\perp}
\sum_1^{\infty}(\frac{(n\pi /L)^2}
{(q_{\perp}^2 + n^2\pi^2/L^2)(q_{\perp}^2 + n^2\pi^2/L^2 + 2gM^2)
})\nonumber ,\\\ea
which, in the large L limit, goes to
\be (e) =  -
\frac{2\theta_0^2}{L^2}\frac{L^{d}}{(2\pi)^{d}}\int
d^{d}p   \frac{p_1^2}{p^2(p^2 + 2gM^2)} = -
\frac{2\theta_0^2}{L^2}\frac{L^{d}}{d(2\pi)^{d}}\int
d^{d}p   \frac{1}{p^2 + 2gM^2}, \ee
from which one finds easily that
\be (e) =\frac{1}{16\pi^2\e}(2gM^2)^{1-\e/2} \theta_0^2 L^{d-2}. \ee
\end{itemize}
Collecting the results (a) to (e) we end up with (\ref{32})  of the
third section.

\end{document}